\documentclass{aa}
\usepackage[varg]{txfonts}
\usepackage{booktabs}
\usepackage{multirow}
\usepackage{float}
\usepackage{color}

\usepackage{natbib}

\def\ps1{Pan-STARRS1}
\def\srg{\textit{SRG}}
\def\art{ART-XC}
\def\ero{eROSITA}
\def\rosat{\textit{ROSAT}}
\def\integral{\textit{INTEGRAL}}
\def\swift{\textit{Swift}}
\def\gaia{\textit{Gaia}}
\def\xmm{\textit{XMM-Newton}}
\def\rxte{\textit{RXTE}}
\def\tess{\textit{TESS}}

\def\sone{SRGA\,J194638.9+704552}
\def\stwo{SRGA\,J204547.8+672642}
\def\sthree{SRGA\,J225412.8+690658}

\def\nh{N_{\rm H}}
\def\nhgal{N_{\rm H,\,Gal}}

\def\doubleline{\vskip 3pt\hrule \vskip 1.5pt \hrule \vskip 5pt}

\def\smfiguresmall#1#2#3{
  \begin{minipage}{0.47\textwidth}
    \begin{minipage}{0.049\columnwidth}
      \rotatebox{90}{\hspace{4.0cm}\footnotesize\phantom{0000}#3}
    \end{minipage}
    \begin{minipage}{0.9\columnwidth}
      \includegraphics[viewport=30 188 556 220,width=0.97\columnwidth]{#1}
      \centerline{\footnotesize #2}
    \end{minipage}
  \end{minipage}
}

\usepackage{amstext}

\begin{document}

\title{Identification of three cataclysmic variables detected by the ART-XC and eROSITA telescopes on board the SRG during the all-sky X-ray survey}

\author{Zaznobin I.\inst{1,2}
  \and Sazonov S.\inst{1,3}
  \and Burenin R.\inst{1,2}
  \and Uskov G.\inst{1}
  \and Semena A.\inst{1}
  \and Gilfanov M.\inst{1,3}
  \and Medvedev P.\inst{1}
  \and Sunyaev R.\inst{1,3}
  \and Eselevich M.\inst{4}
  }

\offprints{Zaznobin I., \email{zaznobin@iki.rssi.ru}}

\institute{Space Research Institute, 84/32 Profsouznaya str., Moscow 117997, Russia
  \and  Sternberg Astronomical Institute, 14 Universitetskij pr., Moscow, Russia
  \and Moscow Institute of Physics and Technology, Institutsky per. 9, 141700 Dolgoprudny, Russia
  \and Max-Planck-Institut f\"{u}r Astrophysik (MPA), Karl-Schwarzschild-Str. 1, D-85741 Garching, Germany
  \and Institute of Solar-Terrestrial Physics, Russian Academy of Sciences, Siberian Branch, Irkutsk, Russia
}

\abstract {We report the discovery of three cataclysmic variables in the data of the first year of the all-sky X-ray survey by the {\it SRG} orbital observatory. The sources were selected for their brightness in the 4--12 keV band in the data of the Mikhail Pavlinsky ART-XC telescope. They are also detected by the eROSITA telescope, which provides accurate localizations and spectral data for a broadband spectral analysis. All three objects were previously known as X-ray sources from the {\it ROSAT} all-sky survey and {\it XMM-Newton} slew survey, but their nature remained unknown. The X-ray spectra obtained by eROSITA and ART-XC are consistent with optically thin thermal emission with a temperature $kT\gtrsim 18$~keV for SRGA\,J194638.9+704552 and SRGA\,J225412.8+690658 and $kT\gtrsim 5$~keV for SRGA\,J204547.8+672642. Together with the inferred high X-ray luminosities ($2\times 10^{32}$ -- $3\times 10^{33}$~erg~s$^{-1}$), this strongly suggests that all three sources are cataclysmic variables (CVs). We have obtained optical photometry and spectroscopy for these objects using the AZT-33IK 1.6 m telescope of the Sayan Observatory. The optical properties confirm the CV nature of the objects. We conclude that SRGA\,J194638.9+704552 is an intermediate polar, SRGA\,J204547.8+672642 is likely a polar or intermediate polar, and SRGA\,J225412.8+690658 is either a magnetic or nonmagnetic CV. We have also measured an orbital period of 2.98~hours for SRGA\,J204547.8+672642 based on \emph{TESS} data. Three out of the planned eight {\it SRG} all-sky surveys have now been completed. We expect to find many new CVs, in particular, magnetic systems, during the survey, and we plan to continue our optical follow-up program. }

\authorrunning{I.\,Zaznobin et al.}

\titlerunning{Identification of cataclysmic variables detected by the SRG}

\maketitle 

\section{Introduction}

The {\it Spectrum-Roentgen-Gamma} ({\it Spektr-RG}, \srg) orbital observatory \citep{Sunyaev_2021} has been carrying out an all-sky X-ray survey since 12 December 2019. The survey is planned to consist of eight consecutive scans of the entire sky, each lasting six months. The \srg\ payload consists of two grazing-incidence telescopes: the extended ROentgen Survey with an Imaging Telescope Array (\ero, \citealt{Predehl_2021}), and the Mikhail Pavlinsky Astronomical Roentgen Telescope -- X-ray Concentrator (\art, \citealt{Pavlinsky_2021_art}). They operate in the overlapping 0.2--8 keV and 4--30 keV energy bands, respectively. The \srg\ survey is unique in a number of ways. In the soft X-ray band (0.2--2.3 keV), \ero\ is expected to detect several million X-ray sources, which is a factor of $\gtrsim\text{ } 30$ more than during the R\"{o}ntgensatellit (\rosat) all-sky survey \citep{Boller_2016} that was conducted 30 years ago. At higher energies, \ero\ and \art\ will survey the whole sky with subarcminute angular resolution for the first time. The \art\ survey in the 4--12 keV energy band is expected to significantly surpass the previous surveys that were carried out in similar (medium X-ray) energy bands in terms of the combination of angular resolution, sensitivity, and uniformity. The \srg\ all-sky survey is expected to have a huge impact on our knowledge of various populations of Galactic and extragalactic objects. Initial results of the mission demonstrate that it meets these expectations \citep{Sunyaev_2021}. 

Recently, a catalog of sources detected in the 4--12 keV band by \art\ during the first year of the all-sky survey (namely, on a combined map of the first and second scans of the sky; ART-XC Sky Surveys 1 and 2, or ARTSS12) has been produced \citep{Pavlinsky_2021_cat} that comprises 867 sources. Most of them ($\sim 750$) are previously known astrophysical objects, the largest group of which are active galactic nuclei (AGN; $\sim 370$), followed by X-ray binaries ($\sim 170$) and cataclysmic variables (CVs; $\sim 100$, including symbiotic systems)\footnote{The approximate sign before the quoted numbers indicates that for some of the objects the suggested classification is currently not reliable.}. 

To place the number of CVs that have been detected by \art\ into context, we note that the published samples of CVs selected in soft X-rays (0.1--2.4~keV) from the \rosat\ all-sky survey comprise $\sim 50$ objects \citep{Schwope_2002,Pretorius_2012,Pretorius_2013}, while the sample of CVs from the \rxte\ slew survey in the 3--20~keV band includes $\sim 25$~objects \citep{Sazonov_2006}. Recently, samples of $\sim 100$ CVs  have been obtained thanks to serendipitous hard X-ray (above 15 keV) surveys carried out by the {\it INTErnational Gamma-Ray Astrophysics Laboratory} (\integral) and \swift\ observatories \citep{Revnivtsev_2008,Oh_2018,Lutovinov_2020}. These data have been accumulated over $\sim 15$~years. It is expected that \art\ will increase the sample of X-ray selected CVs by several times over the course of the four-year survey to up to $\sim 500$ objects. This sample will be of high value for systematic population studies of CVs because the 4--12 keV energy band is ideally suited for selecting CVs of virtually all classes (both magnetic and nonmagnetic, see \citealt{Mukai_2017} for a recent review): It is hardly affected at all by the low-energy (due to photoabsorption) and high-energy cutoffs that are observed in the broadband X-ray spectra of CVs. 

We have started a program of optical follow-up observations of the new and previously unidentified X-ray sources in the ARTSS12 catalog. This has already led to the identification of several AGN, including heavily obscured ones \citep{Zaznobin_2021}. In this paper, we present the results of the identification of three more \art\ sources that have been proved to be CVs. 

\section{Sample}

\begin{table*}
  \caption{Source list.} 
  \label{tab:sources}
  \renewcommand{\tabcolsep}{0.35cm}
  \centering
  \footnotesize
  \begin{tabular}{ccccccc}
    \noalign{\doubleline}   
    & \multispan3\hfil\hspace{2mm} \ero\ position \hfil & \multispan2\hfil \gaia\ position \hfil \\   
    ART-XC source & RA & Dec & $R_{98}$ & RA & Dec & RASS name\\
    \noalign{\hrule}
    \sone & 19 46 38.2 & +70 45 58 & 4.1\arcsec & 19 46 38.09 & +70 45 55.6 & 2RXS\,J$194639.7\!+\!704551$\\
    \stwo & 20 45 48.0 & +67 26 43 & 6.8\arcsec & 20 45 48.04 & +67 26 43.1 & 2RXS\,J$204548.4\!+\!672629$\\
    \sthree & 22 54 13.0 & +69 07 06 & 3.5\arcsec & 22 54 12.99 & +69 07 06.1 & 2RXS\,J$225416.1\!+\!690705$\\
    \noalign{\hrule}
  \end{tabular}
  \tablefoot{$R_{98}$ is the \ero\ source localization radius at 98\% confidence.}
\end{table*}

The objects for this study (see Table~\ref{tab:sources}) were selected from the catalog of point sources detected by the \art\ telescope on the combined map of the first and second sky surveys (12 December 2019 -- 15 December 2020) with a signal-to-noise ratio (S/N) higher than 4.8 in the 4--12 keV energy band. The positions of the sources were measured with an accuracy of better than 30\arcsec\ (at 95\% confidence). All three sources have also been detected by \ero, which allowed us to improve their positions to within a few arcseconds (see Table~\ref{tab:sources}) and to construct their X-ray spectra in a broad energy range from 0.2 to 20 keV (using both \ero\ and \art\ data).

All three objects were discovered as X-ray sources during the \rosat\ all-sky survey (RASS) and were later also detected during the \xmm\ slew survey \citep{Saxton_2008}, but have remained unidentified until now. The improved localizations provided by the \srg\ and the advent of recent optical all-sky surveys (in particular, \gaia), have made it straightforward to identify an optical counterpart for each of these X-ray sources. Table~\ref{tab:sources} provides for each object the \art\ and \ero\ positions, the coordinates of the optical counterpart according to the \gaia\ early data release (EDR3) catalog \citep{Gaia}, and the name of the source in the second \rosat\ all-sky survey (2RXS) source catalogue \citep{Boller_2016}.

\begin{table*}
  \caption{List of X-ray observations.} 
  \label{tab:xray-data}
  \centering
  \footnotesize
\begin{tabular}[t]{llllll}
\toprule
Source &  Dates & Exp.\tablefootmark{a} & $F_{4-12\,{\rm keV}}$\tablefootmark{b} & $F_{0.3-2.2\,{\rm keV}}$\tablefootmark{c} & $\nhgal$\tablefootmark{d}  \\
& & ks & \multicolumn{2}{c}{$10^{-12}$~erg~s$^{-1}$~cm$^{-2}$} & $10^{22}$~cm$^{-2}$ \\
\midrule
\sone & 2020 Jan. 22--24, July 22--24, 2021 Jan. 16--19, 21 & 1.7 & $5.7\pm0.7$ & $2.03^{+0.02}_{-0.03}$  & 0.08\\
\midrule
\stwo\ & 2020 Jan. 3--7, July 1--5, 2021 Jan. 2--6 & 1.8 & $2.3\pm0.5$ & $0.77^{+0.02}_{-0.03}$ & 0.14 \\
\midrule
\sthree\ & 2020 Jan. 27--30, July 28--31, 2021 Jan. 21--23 & 1.1 & $1.9\pm0.7$ & $0.25^{+0.01}_{-0.01}$ & 0.41\\
\bottomrule
\end{tabular}
\tablefoot{
    \tablefoottext{a}{Total \art\ exposure on the source.} \\
    \tablefoottext{b}{\art\ flux in the 4--12 keV energy band averaged over the three sky surveys, estimated assuming a Crab-like spectrum.} \\
    \tablefoottext{c}{\ero\ flux in the 0.3--2.2 keV energy band averaged over the three sky surveys, estimated using the best-fit absorbed power-law spectral models from Table~\ref{tab:xray_table}.}\\
    \tablefoottext{c}{Galactic HI column density from \citet{BEKHTI2016}.}\\
    The flux errors correspond to the 68\% confidence level.
}
\end{table*}

\section{X-ray observations}

All three objects have by now been observed during the first three \srg\ all-sky surveys. As a result of their high ecliptic latitudes ($b_{\rm ecl}=79.5^\circ$, $74.1^\circ$, and $63.8^\circ$ for \object{\sone}, \object{\stwo}, and \object{\sthree}, respectively) and the strategy of the \srg\ survey (namely, scanning great circles that intersect at the ecliptic poles; see \citealt{Sunyaev_2021}), all three sources received relatively long observing times. Specifically, the \srg\ visited them over periods of 3--5~days during each all-sky survey, compared to $\sim 1$~day for typical sources at low ecliptic latitudes. However, in reality, each of these visits consisted of a few dozen short ($\sim 20$~s and $\sim 40$~s for \art\ and \ero, respectively) passages separated by 4~hours, so that the total accumulated exposure times are just $\sim 1$~ks for \art\ and a factor of $\sim 4$ longer for \ero\footnote{The fields of view of \art\ and \ero\ are 36 arcmin and 1 deg, respectively.} (see Table~\ref{tab:xray-data}).

In what follows, we focus on the fluxes and spectra of the sources averaged over all available \srg\ observations. Table~\ref{tab:xray-data} provides the average fluxes in the 0.3--2.2 and 4--12 keV energy bands, where the maximum sensitivity is achieved by \ero\ and \art, respectively \citep{Sunyaev_2021}.

We also provide in Table~\ref{tab:xray-data} the column densities of neutral hydrogen throughout the Galaxy, $\nhgal$, in the direction of the objects \citep{BEKHTI2016}. Because the distances to the sources are all larger than $\sim 700$~pc and they are located relatively high above the Galactic plane ($b= 21.1^{\circ}$, $14.9^{\circ}$, and $8.6^{\circ}$ for \sone, \stwo, and \sthree, respectively), we may expect the bulk of the Galactic HI column to be located between us and the objects. This issue is discussed in more detail in Section~\ref{s:optical}. 

\subsection{Spectral analysis}

\begin{figure}
\centering  
\includegraphics[width=0.9\columnwidth]{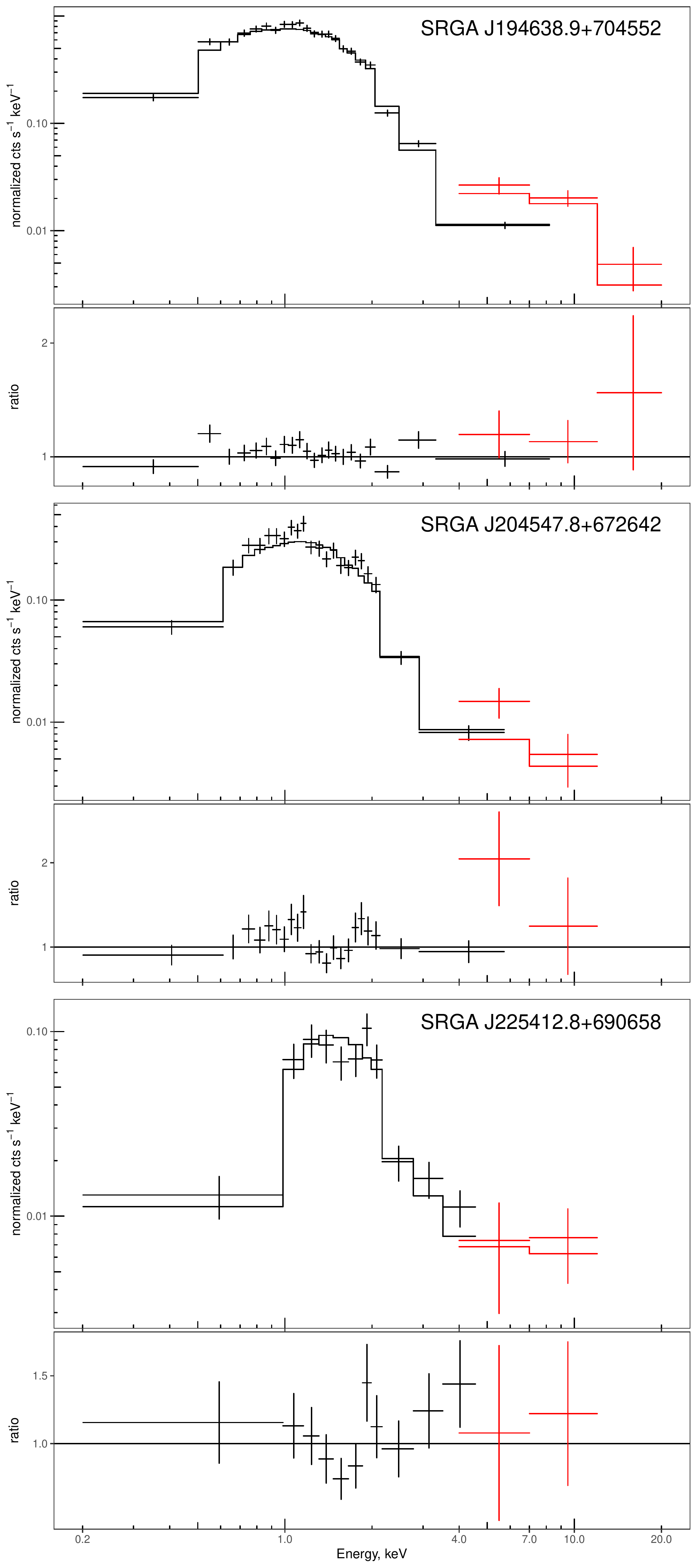}
\caption{X-ray spectra (upper panels) obtained by \ero\ (black) and \art\ (red) and their best-fit \textsc{phabs*bremss} models. The lower panels show the corresponding data-to-model ratio plots. The horizontal line indicates a ratio of 1.
}
\label{fig_sim}
\end{figure}

\begin{table}
\caption{X-ray spectral properties.} 
\label{tab:xray_table}
\renewcommand{\arraystretch}{1.5}
\renewcommand{\tabcolsep}{0.1cm}
\centering
\footnotesize
\begin{tabular}[t]{rccc}
\toprule
Parameter & \multicolumn{3}{c}{SRGA source} \\
& 194638.9+704552 & 204547.8+672642 & 225412.8+690658 \\
\midrule
\addlinespace[0.3em]
\multicolumn{4}{c}{\emph{POWER LAW}}\\
$\nh$\tablefootmark{a} & $0.12^{+0.02}_{-0.02}$ & $0.25^{+0.04}_{-0.06}$ & $0.50^{+0.24}_{-0.13}$\\
$\Gamma$ & $1.3^{+0.1}_{-0.1}$ & $1.6^{+0.2}_{-0.2}$ & $0.8^{+0.3}_{-0.3}$\\
norm & $9.8^{+0.6}_{-0.6}\times 10^{-4}$ & $5.0^{+0.9}_{-0.7}\times 10^{-4}$ & $1.8^{+0.8}_{-0.5}\times 10^{-4}$\\
$\chi^2$ & 182.9 & 77 & 7.7\\
dof\tablefootmark{c} & 167 & 47 & \vphantom{2} 10\\
\midrule
\addlinespace[0.3em]
\multicolumn{4}{c}{\emph{BREMSSTRAHLUNG}}\\
$\nh$ & $0.11^{+0.01}_{-0.01}$ & $0.19^{+0.03}_{-0.03}$ & $0.69^{+0.17}_{-0.14}$\\
$kT$\tablefootmark{b} & $ 40^{+53}_{-17}$ & $ 10^{+15}_{-4}$ & $>21$\\
norm & $1.8^{+0.4}_{-0.2}\times 10^{-3}$ & $6.3^{+0.8}_{-0.4}\times 10^{-4}$ & $7.9^{+0.8}_{-3.2}\times 10^{-4}$\\
$\chi^2$ & 182.9 & 77.0 & 10.7\\
dof & 167 & 47 & \vphantom{1} 10\\
\midrule
\addlinespace[0.3em]
\multicolumn{4}{c}{\emph{APEC}}\\
$\nh$ & $0.11^{+0.01}_{-0.01}$ & $0.20^{+0.03}_{-0.02}$ & $0.71^{+0.24}_{-0.11}$\\
$kT$ & $ >18$ & $  7.1^{+   3.8}_{-  1.7}$ & $ >18$\\
norm & $5.1^{+0.9}_{-0.7}\times 10^{-3}$ & $1.65^{+0.13}_{-0.12}\times 10^{-3}$ & $1.7^{+0.2}_{-0.4}\times 10^{-3}$\\
$\chi^2$ & 184.6 & 72.2 & 11.9\\
dof & 167 & 47 & 10\\
\bottomrule
\end{tabular}
\tablefoot{
\tablefoottext{a}{Hydrogen column, in units of $10^{22}$~cm$^{-2}$}\\
\tablefoottext{b}{Temperature in keV.}\\
\tablefoottext{c}{Degrees of freedom.}\\
  The errors and limits correspond to the 90\% confidence level.
  }
\end{table}

Based on the \ero\ and \art\ data, we performed an X-ray spectral analysis in the 0.2--12~keV energy range for \stwo\ and \sthree\ and in the 0.2--20~keV range for \sone. The \ero\ raw data were processed by the calibration pipeline at the Space Research Institute (IKI, Moscow), which was built using the components of the \ero\ Science Analysis Software System (eSASS). 

We extracted the source spectra and light curves using a circular region with a radius of 60 arcsec centered on the source position. An annulus region with the inner and outer radii of 150 and 300~arcsec around each source was used for background extraction. We masked out the circular regions of 40~arcsec radius around all sources that overlap with the studied sources and background regions and detected with $S/N>4$. The \ero\ spectra were binned to have at least 25 counts in each spectral interval. 

We processed the \art\ data with the analysis software \textsc{artproducts} v0.9 \citep{Pavlinsky_2021_art} using the calibration files version 20200401. We tried to extract counts in the 4--7, 7--12, and 12--20~keV energy intervals for all three sources. However, \stwo\ and \sthree\ have very few counts in the 12--20 keV bin. Therefore, we only used the 4--7 and 7--12~keV channels in the spectral analysis of these two sources.

We performed the spectral analysis using \textsc{xspec}\footnote{https://heasarc.gsfc.nasa.gov/xanadu/xspec/} v12.11.0. \citep{xspecArnaud1996} and assessed the quality of spectral fits using a $\chi^2$ statistic. For some of the spectral models discussed below, the parameter uncertainties were estimated using Monte Carlo Markov chains (MCMC) based on the Goodman--Weare algorithm \citep{GOODMAN} with $15\times10^3$ steps (namely for \sthree\ in the case of \emph{apec}). No cross-normalization constant between the \ero\ and \art\ data is used below because introducing this coefficient does not lead to a significant improvement in the spectral fit quality. 

\subsection{Models}

We first tried to describe the spectra by a power-law model modified by photoabsorption (\emph{phabs*powerlaw} model in \textsc{xspec}) and obtained satisfactory fits in all three cases (Table~\ref{tab:xray_table}). The inferred absorption columns along the line of sight, $\nh$, are marginally higher than the total Galactic columns of neutral hydrogen in the direction of \sone\ and \stwo,\, whereas for \sthree,\ $\nh$ is consistent with $\nhgal$. All three spectra are hard. The inferred photon indices $\Gamma< 1.8$ are suggestive of thermal emission of an optically thin plasma and typical of CVs, both magnetic and nonmagnetic CVs (e.g., \citealt{Byckling_2010,Burenin_2016}). 

To investigate this hypothesis further, we fitted the spectra with an absorbed thermal bremsstrahlung model (\emph{phabs*bremss}). This resulted in nearly the same fit quality (see Table~\ref{tab:xray_table}). We can place reliable lower limits on the temperature of the plasma of $kT\gtrsim 20$~keV for \sone\ and \sthree\ and $kT\gtrsim 5$~keV for \stwo. 
Figure~\ref{fig_sim} shows the best fits for this model. 

We next tried to describe the spectra by the \emph{apec} model for a collisionally ionized diffuse gas. As expected, this did not lead to a significant improvement in the fit quality compared to \emph{bremss} for \sone\ and \sthree\ because for such high temperatures as inferred here the plasma is expected to be nearly fully ionized, with no strong emission lines produced (except for weak lines of H- and He-like Fe and Ni near 7~keV, which include too few photons in the spectra to make a difference). For these sources, the lower limits on the plasma temperature inferred from the \emph{apec} model, $kT\gtrsim 18$~keV, are somewhat less stringent than the corresponding values for \emph{bremss}. For \stwo, \emph{apec} provides a minor improvement in the fit quality over \emph{bremss} and bounds the plasma temperature within a relatively narrow interval around $kT\sim 7$~keV. This indicates the presence of non-negligible line emission in the \ero\ spectrum of this source. 

In summary, both the estimated X-ray luminosities ($\sim 10^{32}$--$10^{33}$~erg~s$^{-1}$) and the X-ray spectra being consistent with emission from hot, optically thin plasma strongly indicate that all three sources are CVs. 

\section{Optical observations}
\label{s:optical}

\begin{figure*}
  \centering
    \sone \hspace{1cm} \stwo \hspace{1cm} \sthree\\
    \includegraphics[width=0.52\columnwidth]{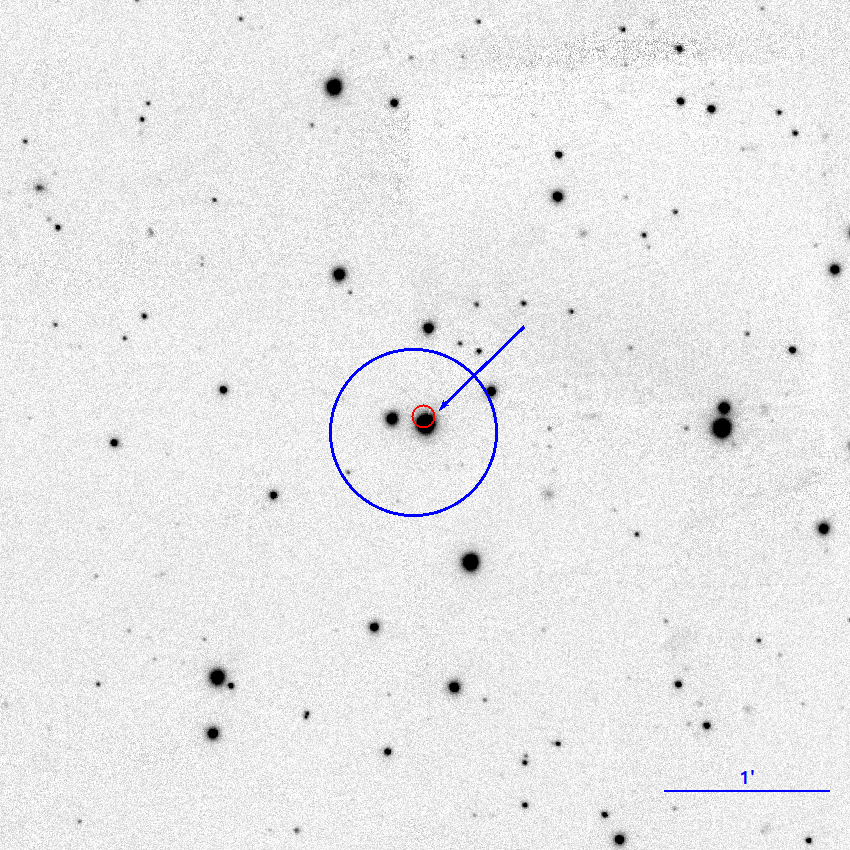}
    \includegraphics[width=0.52\columnwidth]{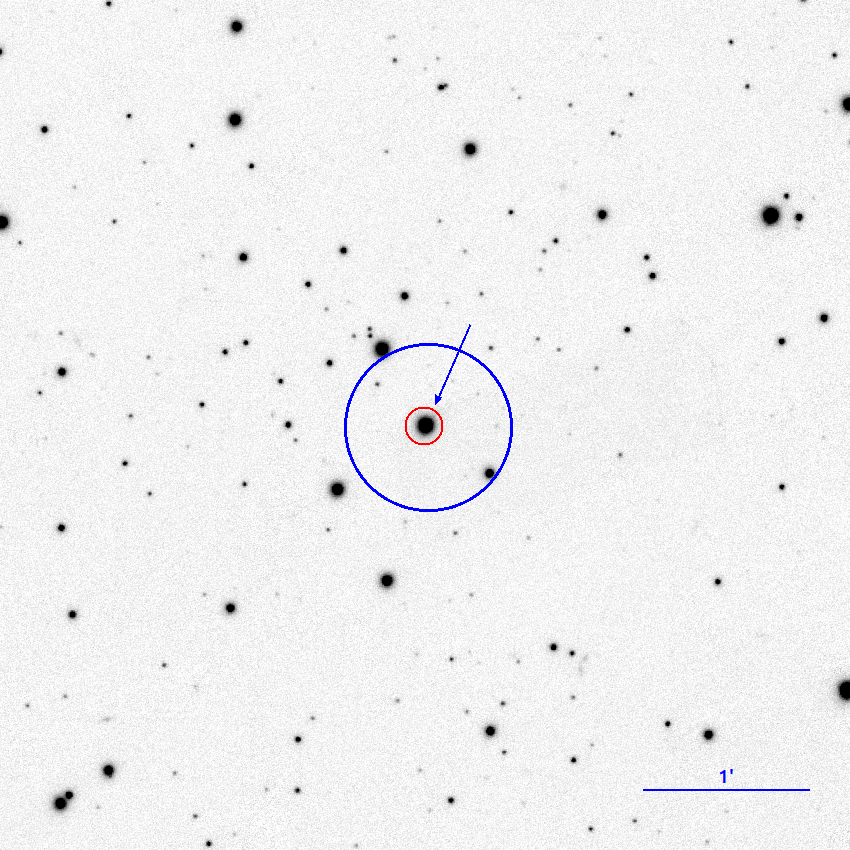}
    \includegraphics[width=0.52\columnwidth]{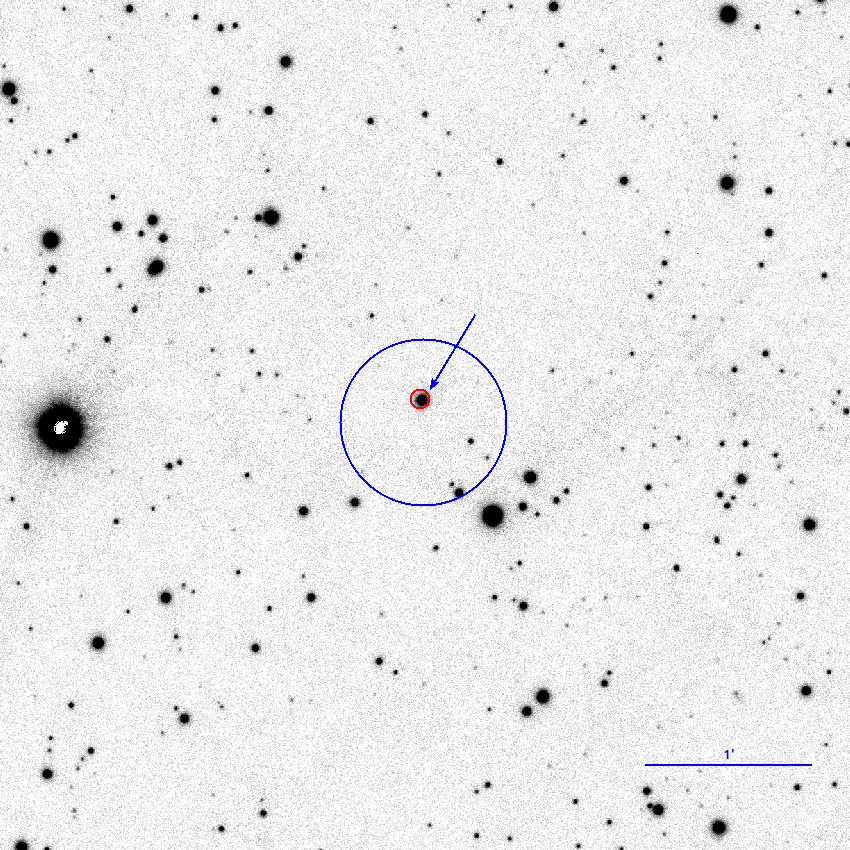}
  \caption{\ps1\ images in the \emph{r} filter around \sone\ (left), \stwo\ (middle), and \sthree\ (right). The size of each image is $5\arcmin \times 5\arcmin$. The blue circle with a radius of 30\arcsec\ indicates the \art\ location. The red circle shows the \ero\ 98\% location region. The optical counterpart is indicated by the arrow.}
  \label{fig:images}
\end{figure*}

For each of the three X-ray sources, the archival optical images show a single relatively bright star within the \ero\ location region (Fig.~\ref{fig:images}). These stars are known to be variable from previous observations, which corroborates that these are CVs.

Some key optical characteristics of these objects are listed in Table~\ref{tab:mag} and are discussed in Section~\ref{s:results} below. Specifically, for each star, we provide the parallax distance based on \cite{Gaia}, the apparent visual magnitude and the amplitude of its variability, the interstellar extinction to the object, and the absolute visual magnitude. Our estimates of the extinction are based on (i) the \cite{dust98} dust map (SFD) recalibrated according to \cite{Schlafly_11}, which provides the total Galactic extinction in a given direction\footnote{https://irsa.ipac.caltech.edu/applications/DUST/}, and (ii) the Bayestar19 model of the three-dimensional dust distribution in the Galaxy \citep{Bayestar19}, which allows the distance to the object to be taken into account\footnote{http://argonaut.skymaps.info/}. We adopt $R_V=3.1$ and $R_{g-r} = E(g-r)/E(B-V)=0.98 \pm 0.02$ \citep{Schlafly_11} to convert $E(B-V)$ and $E(g-r)$ into visual extinction, $A_V$. The SFD and Bayestar19 estimates of $A_V$ prove to not differ significantly from each other for the stars under consideration, which is expected because of their large distances and fairly high altitudes above the Galactic plane.

\begin{table}
  \caption{Optical properties of the CVs.} 
  \label{tab:mag}
  \centering
  \begin{tabular}{lccccc}
    \noalign{\doubleline}   
    Object & $D$\tablefootmark{a}, pc & $V$ & $\Delta V$\tablefootmark{b}, & $A_V$\tablefootmark{c}, & $M_V$ \\
    \noalign{\hrule}
    \sone & $1816\pm 43$ & 14.16 & $\pm0.23$ & 0.5 & 2.3 \\
    \stwo & $848\pm 9$ & 13.83 & $\pm0.34$ & 0.9 & 3.3 \\
    \sthree & $686\pm 12$ & 16.46 & $\pm2.11$ & 2.4 & 4.9 \\
    \noalign{\hrule}
  \end{tabular}
  \tablefoot{
  \tablefoottext{a}{Distance based on \cite{Gaia}.}\\
  \tablefoottext{b}{Variability amplitude based on ZTF and ASAS-SN data.}\\
  \tablefoottext{c}{Extinction based on \cite{dust98,Schlafly_11}, and \cite{Bayestar19}.}
   }
\end{table}

\subsection{Spectroscopy}

To study these objects in detail, we conducted follow-up optical observations with the Sayan Observatory 1.6 m AZT-33IK telescope. For spectroscopy, we used the ADAM low- and medium-resolution spectrograph \citep{adam,adam16}. The observations were made using the VPHG600G volume-phase holographic grating, which provides the 3650\AA\ -- 7250\AA\ spectral range and $\sim 9\AA$ spectral resolution with a 2\arcsec\ slit. The typical image quality was about $1\arcsec\!.8$--$2\arcsec$. During the observations of each object, halogen and He-Ne-Ar lamp calibration spectra were obtained after a series of spectroscopic images. After each exposure, we shifted the position of the object along the slit by $10\arcsec$--$15\arcsec$ in a random direction to reduce the interference fringes in the spectra. We processed the data using the \texttt{IRAF}\footnote{http://iraf.noao.edu} software package and our own software. Spectral fluxes were  calibrated using observations of spectrophotometric standards from the ESO list\footnote{https://www.eso.org/sci/observing/tools/standards/spectra.html}.

To estimate the intrinsic width of emission lines, we quadratically subtracted the instrument line broadening from the line widths measured from the observed spectra. The correction of the spectra for extinction was made using the \textsc{IRAF} \emph{deredden} task. The extinction for each object was estimated as described below.

\begin{table*}
  \caption{Log of optical observations with the AZT-33IK telescope.} 
  \label{tab:obs}
  \centering
  \begin{tabular}{llcccccc}
    \noalign{\doubleline}   
    & \multispan4\hfil Spectroscopy\hfil & \multispan3\hfil Photometry\hfil\\   
    Object & Date & Exp. time & Grism & Slit & Date & Exp. time & Filter\\
    \noalign{\hrule}
    \sone & 2020-10-23 & $3 \times 60$~s & VPHG600G & $2\arcsec$ & 2021-06-17 & $420 \times 15$~s & r \\ 
    \stwo & 2020-10-21 & $3 \times 200$~s & VPHG600G & $2\arcsec$ & 2021-06-18 & $540 \times 15$~s & r \\ 
    \sthree & 2021-05-14 & $5 \times 100$~s & VPHG600G & $2\arcsec$ & 2021-06-19 & $240 \times 30$~s & r \\ 
    \noalign{\hrule}
  \end{tabular}
\tablefoot{
  The exp. time columns show the number of obtained images during continuous observations and the exposure time of each frame. The date columns show the observation dates in the YYYY-MM-DD format, UTC+0.
}
\end{table*}

\subsection{Photometry}

During the period from 17 to 20 June 2021, we conducted photometric observations of the objects using the CCD photometer of the AZT-33IK telescope. It consists of a focal reducer and the Andor iKon-M 934 camera with the BD\_DD $1024\times1024$ detector, which provide a $6.3\arcmin \times 6.3\arcmin$ field of view with a $0\arcsec\!.372$ pixel scale. For each object, the observations were carried out continuously for 2 hours or more. The quality of direct images was not worse than $1\arcsec.7$ in the SDSS $r$ filter.

We processed the images in a standard way using \textsc{iraf} and our own software. Aperture photometry was made using the \emph{apphot} task from the \textsc{iraf} \emph{digiphot} package. The source fluxes were measured relative to nearby bright stars, with the aperture size for each observation series defined such that the maximum possible S/N was obtained.

We calibrated the measured magnitudes using the magnitudes of secondary photometric standards in the source field with a brightness comparable to or greater than the flux of the CV candidates. We used point-spread-function magnitudes from \ps1\ DR2 \citep{ps1} for calibration. The processing of light curve data was performed using the \emph{Python Lightkurve} package \citep{Lightkurve}. Further details of the optical observations are given in Table~\ref{tab:obs}.

\subsection{Results}
\label{s:results}

\begin{figure*}
  \centering
  \vfill
  \sone
  \vfill
  \smfiguresmall{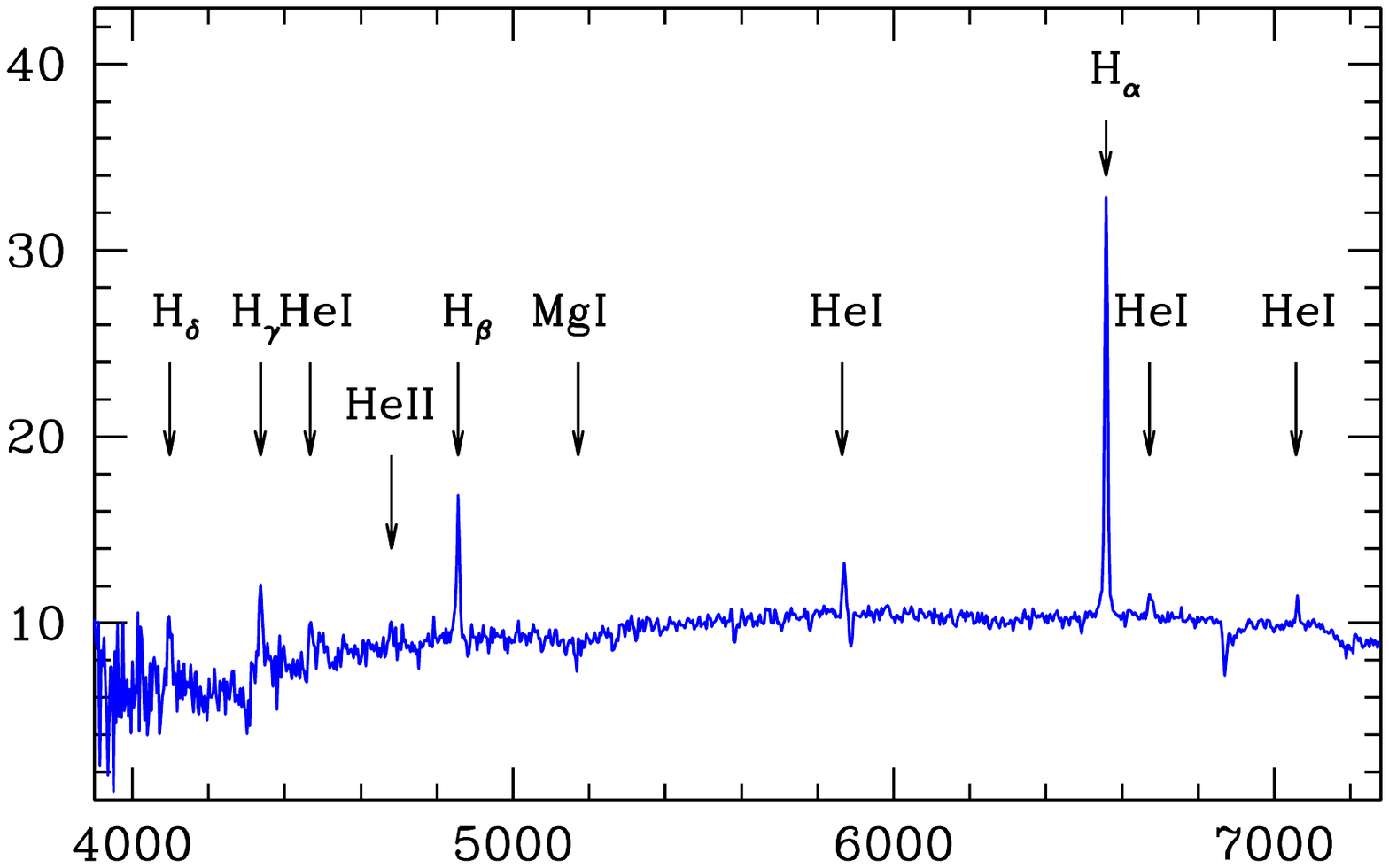}{$\lambda$, \AA}{flux, $10^{-15}$~erg\,s$^{-1}$\,cm$^{-2}$\,$\AA^{-1}$}
  \includegraphics[width=0.78\columnwidth]{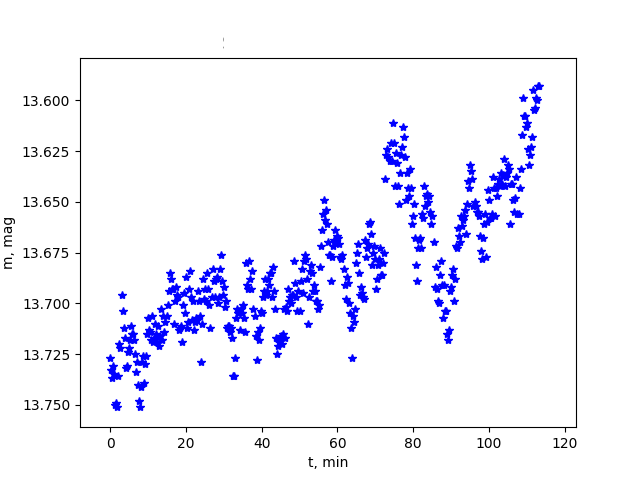}
  \vfill
  \vspace{-3.5cm}
  \stwo
  \vfill
   \smfiguresmall{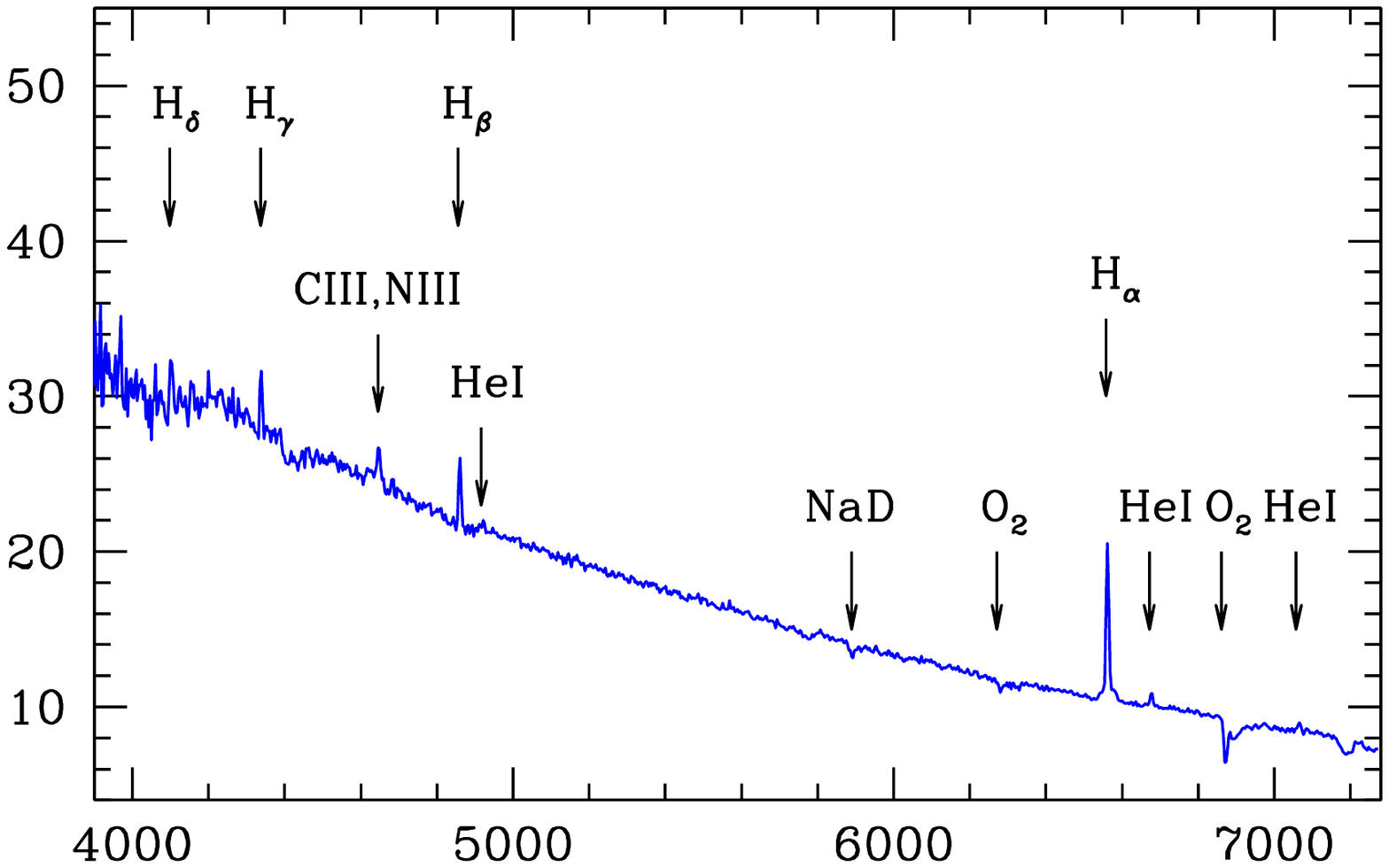}{$\lambda$, \AA}{flux, $10^{-15}$~erg\,s$^{-1}$\,cm$^{-2}$\,$\AA^{-1}$}
  \includegraphics[width=0.8\columnwidth]{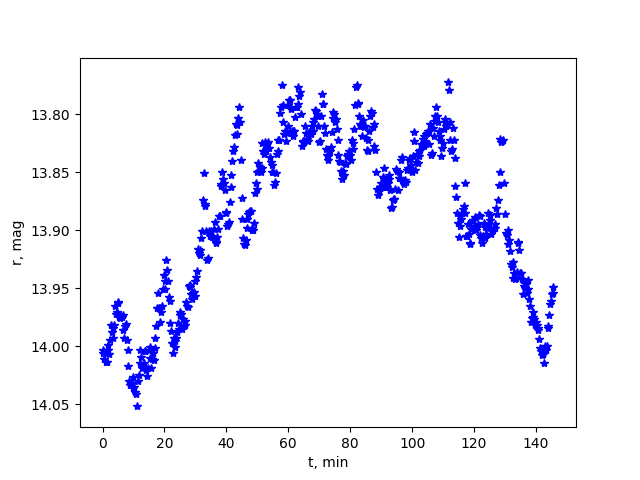}
  \vfill
  \vspace{-3.5cm}
  \sthree
  \vfill
  \smfiguresmall{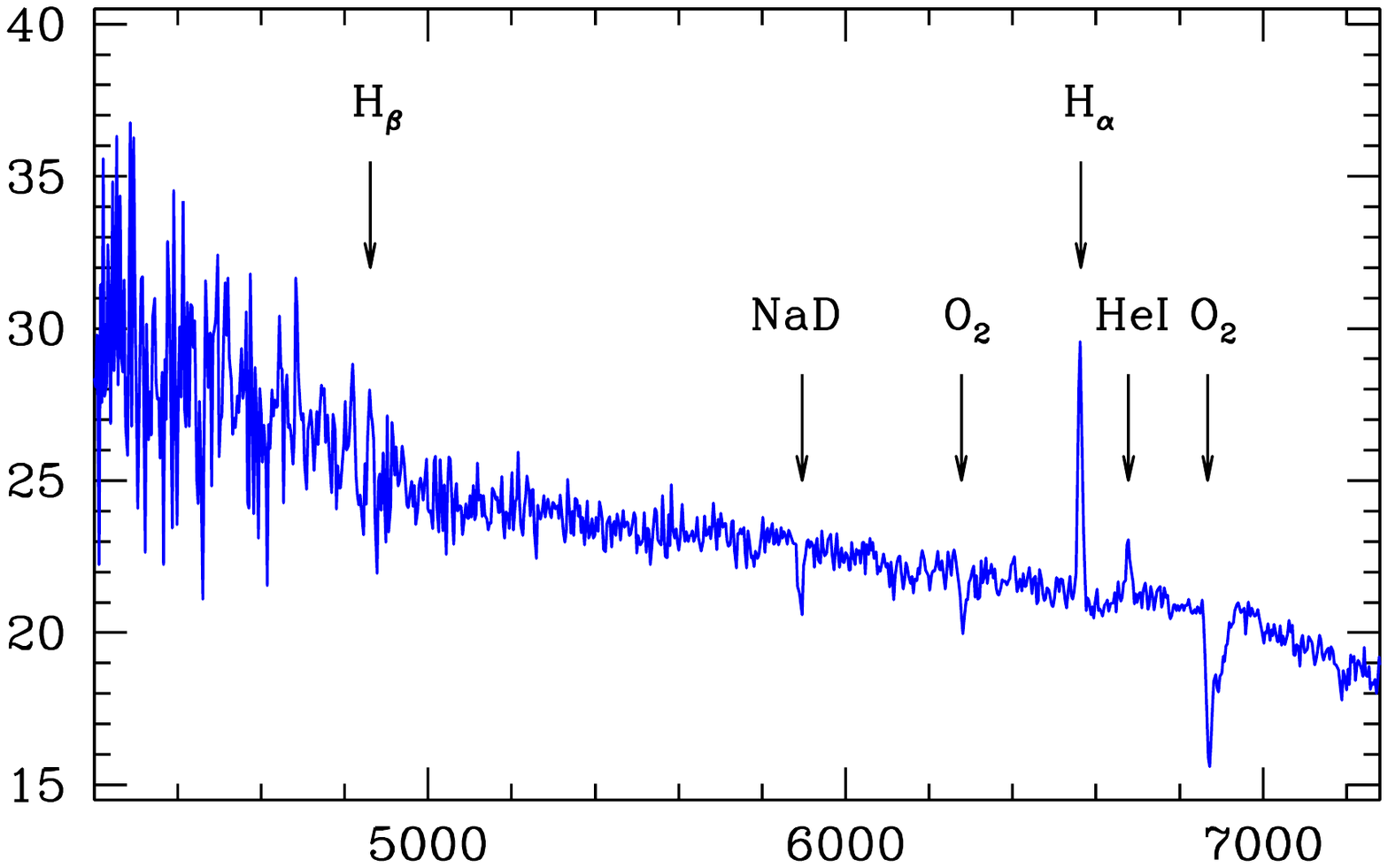}{$\lambda$, \AA}{flux, $10^{-15}$~erg\,s$^{-1}$\,cm$^{-2}$\,$\AA^{-1}$}
  \includegraphics[width=0.8\columnwidth]{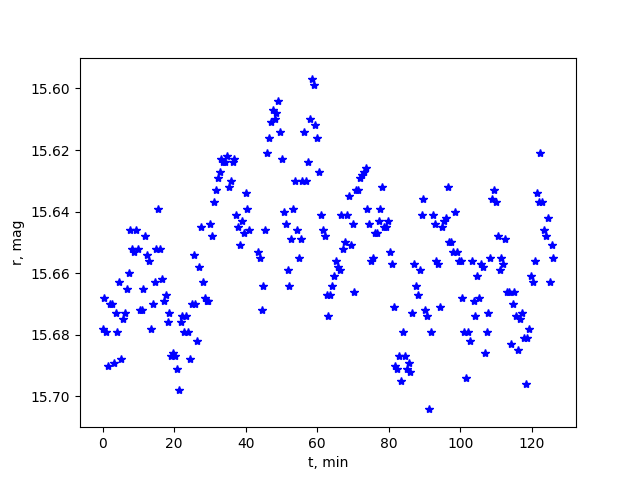}
  \vspace{-3.5cm}
  \caption{Optical spectra (left) and light curves (right) of the objects. The spectra are corrected for interstellar extinction, and prominent spectral features are indicated. The light curves are obtained in the SDSS \emph{r} filter, and the time is measured from the beginning of the observation.}
  \label{fig:spec_lc}
\end{figure*}

Figure~\ref{fig:spec_lc} shows the measured optical spectra and light curves. Table~\ref{tab:spec} presents the characteristics of the emission lines detected in the spectra. We now discuss the results of archival and follow-up optical observations on a source-by-source basis.

\begin{table}
  \caption{Optical emission lines.} 
  \label{tab:spec}
  \renewcommand{\tabcolsep}{0.08cm}
  \centering
  \footnotesize
  \begin{tabular}{lcccc}
    \noalign{\doubleline}
    Lines & Wavelength\tablefootmark{a} & Flux\tablefootmark{b} & Eq. width\tablefootmark{c} & FWHM\\
     & \AA & $10^{-14}$ erg~s$^{-1}$~cm$^{-2}$ & \AA & km s$^{-1}$ \\
    \noalign{\hrule} 
    \multicolumn{5}{c}{\sone}\\
    H$_\delta$ & 4098.2 & $<2.6$ & -- & --\\
    H$_\gamma$ & 4337.2 & $2.5^{+1.4}_{-0.5}$ & $-6.0^{+5.5}_{-1.4}$ & $337\pm 57$\\ 
    HeI$\lambda4471$ & 4468 & $<2.9$ & -- & --\\
    HeII$\lambda4686$ & 4680 & $<2.1$ & -- & --\\
    H$_\beta$ & 4856.7 & $4.3\pm 0.5$ & $-8.2\pm 0.6$ & $238\pm 48$\\ 
    HeI$\lambda 5876$ & 5870.4 & $2.1^{+0.7}_{-0.4}$ & $-3.2^{+0.9}_{-0.6}$ & $336\pm 73$\\ 
    H$_\alpha$ & 6558.6 & $14.8\pm 0.8$ & $-20.8\pm 2.1$ & $205\pm 36$\\ 
    HeI$\lambda 6678$ & 6675.1 & $<1.5$ & -- & --\\
    HeI$\lambda 7065$ & 7060.1 & $<1.1$ & -- & --\\
    \multicolumn{5}{c}{\stwo}\\
    H$_\delta$ & 4101.9 & $<1.6$ & -- & --\\
    H$_\gamma$ & 4338.2 & $1.1^{+0.5}_{-0.2}$ & $-1.2^{+0.5}_{-0.2}$ & $240\pm 90$\\ 
    CIII,\,NIII & & & & \\
    $\lambda\lambda 4640,\,4650$ & 4647.4 & $1.2^{+0.8}_{-0.3}$ & -- & --\\
    HeII$\lambda4686$ & 4684 & $<0.5$ & -- & --\\
    H$_\beta$ & 4860.0 & $1.4\pm 0.4$ & $-1.7\pm 0.5$ & $156\pm 64$\\ 
    HeI$\lambda 4922$ & 4922.3 & $<0.6$ & -- & --\\
    H$_\alpha$ & 6562.1 & $4.9\pm 0.3$ & $-9.2\pm 0.9$ & $100\pm 23$\\ 
    HeI$\lambda 6678$ & 6678.5 & $<0.6$ & -- & --\\
    HeI$\lambda 7065$ & 7065.5 & $<0.6$ & -- & --\\
    \multicolumn{5}{c}{\sthree}\\
    H$_\beta$ & 4862.6 & $<7.9$ & -- & --\\ 
    H$_\alpha$ & 6561.5 & $16.9\pm 2.0$ & $-5.0\pm 0.7$ & $188\pm 42$\\ 
    HeI$\lambda 6678$ & 6677.1 & $4.3\pm 1.1$ & $-1.2\pm 0.3$ & $222\pm 141$\\ 
    \noalign{\hrule}
  \end{tabular}
\tablefoot{
\tablefoottext{a}{Measured line wavelength.}\\
\tablefoottext{b}{The fluxes are not corrected for the interstellar extinction. Upper limits are provided for lines with an S/N lower than $2\sigma$.}\\
\tablefoottext{c}{Negative values correspond to emission lines. Uncertainties correspond to the $1\sigma$ (68\%) confidence interval.}
}
\end{table}

\subsubsection{\sone}

This X-ray source is identified with the source ID 2263241129823231104 in the \gaia\ EDR3 catalog. According to the UCAC4 catalog \citep{ucac4}, it has average magnitudes $B=15.00$ and $V=14.16$. The object has also been observed in the Zwicky Transient Facility (ZTF\footnote{https://www.ztf.caltech.edu/}) survey with the Palomar Observatory 48-inch Samuel Oschin Telescope. The transient ZTF18absajkj\footnote{https://alerce.online/object/ZTF18absajkj} was discovered on 31 August 2018, and was rediscovered as transient ZTF19acyzjtl on 25 November 2019. According to the Automatic Learning for the Rapid Classification of Events (ALeRCE\footnote{http://alerce.science/}) \citep{alerce1, alerce2}, it is more likely to be a CV than a cepheid or eclipsed binary star. Its mean magnitudes in the \emph{ZTF} filters \citep{ZTF} are $g_{mean} = 14.54$ and $r_{mean} = 13.74$. Based on the ZTF light curve, which spans $\sim 3$~years, we can estimate the amplitude of the source variability at $\Delta V= \pm 0.23$. 

The spectrum of the object (see Fig.~\ref{fig:spec_lc}) is typical of CVs (e.g., \citealt{Warner}). Specifically, it shows a series of hydrogen and helium emission lines. Some Fraunhofer absorption lines are also visible. The Balmer decrement ${\rm H}_{\alpha}/{\rm H}_{\beta} = 3.4 \pm 0.4$ (hereafter, the quoted uncertainties are at $1\sigma$ confidence level, unless noted otherwise), which indicates a moderate extinction if an intrinsic ratio ${\rm H}_{\alpha}/{\rm H}_{\beta}=2.86$ is adopted \citep{Osterbrock}. Using the $E(B-V) = 1.97\log_{10} [({\rm H}_{\alpha}/{\rm H}_{\beta})_{\rm obs}/2.86]$ relation from \cite{Osterbrock}, we obtain $E(B-V) = 0.15 \pm 0.10$, which is consistent with the interstellar extinction to the source.

During the 114 minutes of AZT-33IK photometry, we obtained 15 images with an exposure of $420$~s. Over this period, the reference star magnitude varied by $0.008^m$ (root-mean square), while \sone\ exhibited much stronger variations (see Fig.~\ref{fig:spec_lc}). Because the intervals between the individual exposures were not strictly constant, we used a Lomb--Scargle periodogram \citep{Lomb,scargle} for the timing analysis. The derived power spectrum (not shown) reveals strong red-noise-like variability (already obvious from the high-amplitude stochastic fluctuations in the light curve) modified by a window function. There are no signs of periodicity or quasi-periodicity. Due to the window function, it is difficult to place upper limits on the amplitude of any periodic signal. However, the general rising trend evident in the light curve suggests that the orbital period of this CV is longer than 2~hours.

If this object were a magnetic CV, we might expect to see signs of the white dwarf spin period in its light curve, which are usually in the range of $\sim 1$~min to $\sim 1$~hour for this class of objects (e.g., \citealt{Mukai_2017}). However, spin modulation is not always evident in the optical light of magnetic CVs, thus its absence in the optical light curve of \sone\ does not rule out a polar or  intermediate polar origin. 

\subsubsection{\stwo}

\begin{figure*}
  \centering
  \vfill
  SRGA\,J$204547.8\!+\!672642$
  \vfill
    \includegraphics[width=0.97\columnwidth]{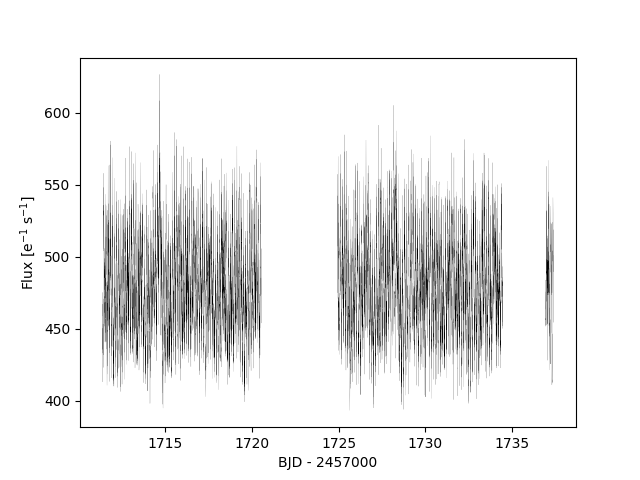}   
    \includegraphics[width=0.97\columnwidth]{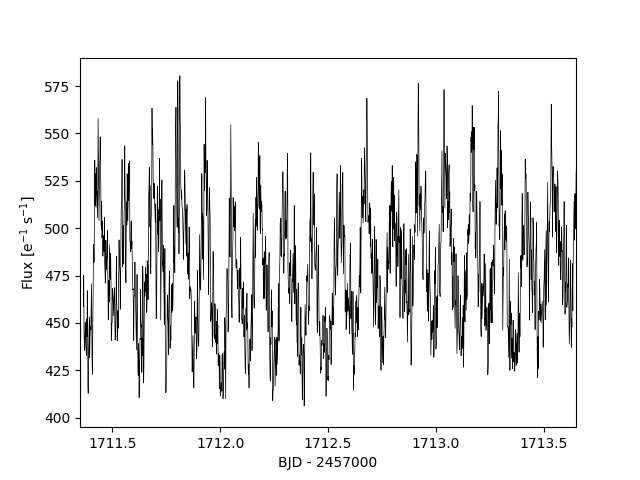}   
    \includegraphics[width=0.97\columnwidth]{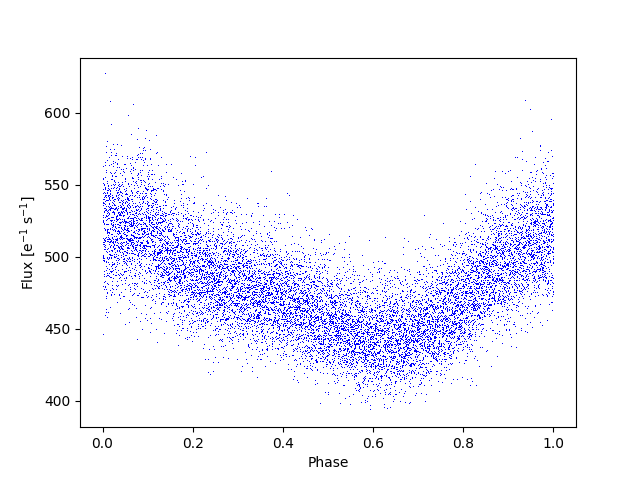}
    \includegraphics[width=0.97\columnwidth]{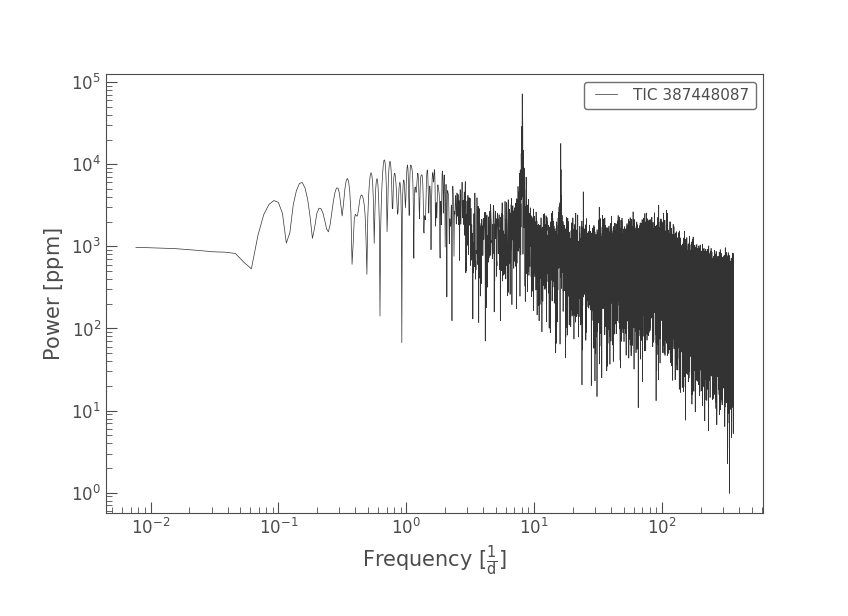}
  \caption{\tess\ observations of \stwo. Top row: light curves obtained in sector 15 (full data set on the left, and the first two days of observations on the right). Bottom left: Light curve folded on the 2.98-hour period. Bottom right: Lomb--Scargle periodogram.}
  \label{fig:tess2045}
\end{figure*}

\begin{figure}
    \centering  
    \includegraphics[width=1\columnwidth]{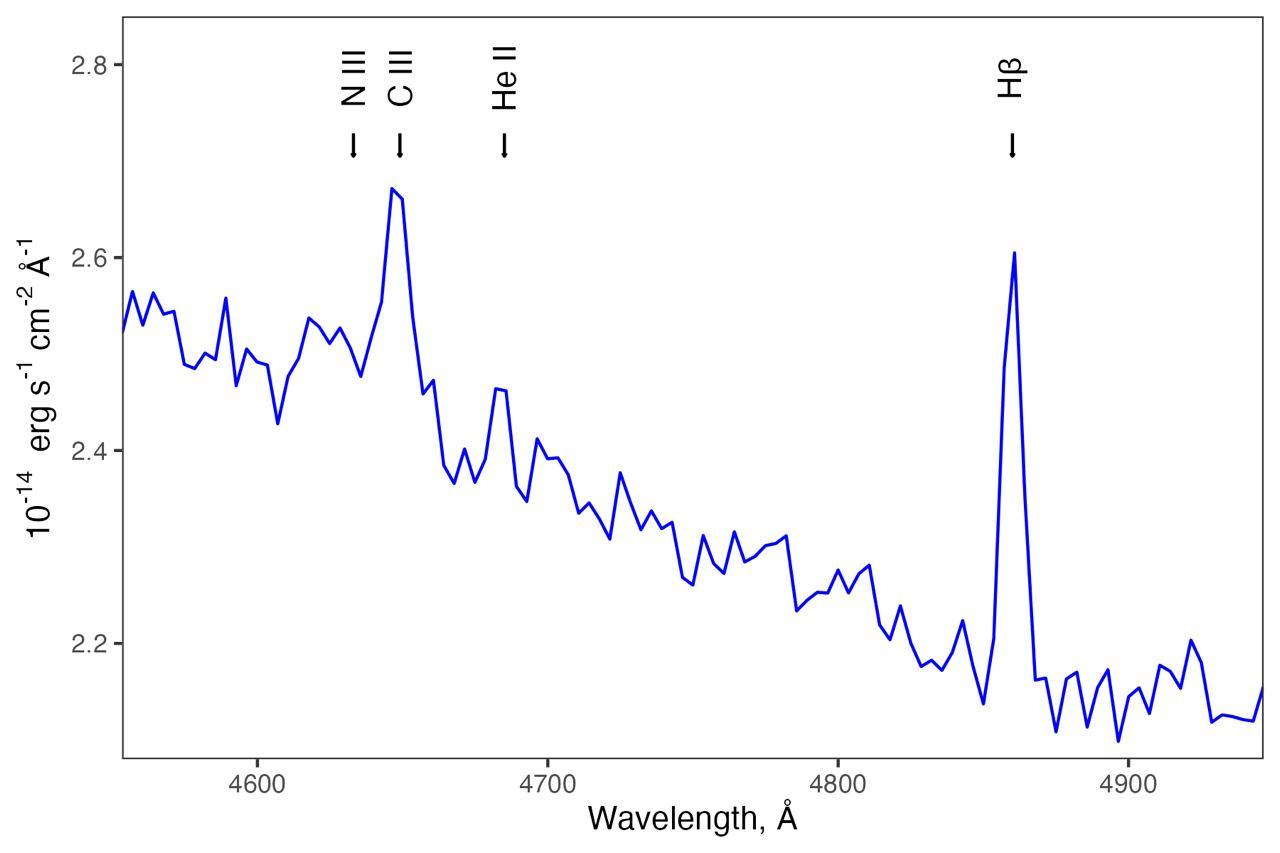}
    \caption{Zoom into the 4550--4950~\AA\ range of the optical spectrum of \stwo. The Bowen blend and H$_{\beta}$ are visible, while HeII$\lambda$4686 is not significantly detected.}
    \label{Bowen:blend}
\end{figure}

This X-ray source is identified with the nonperiodic variable star ASASSN-V~J204548.04+672643.0\footnote{https://asas-sn.osu.edu/variables/AP15943107} from the All Sky Automated Survey for SuperNovae (ASAS-SN) catalog of variable stars VI \citep{asas20}, with an average $V$ magnitude $V_{mean}=13.83$ and an amplitude of variations $\Delta V=0.34$. In the \gaia\ EDR3 catalog, the object is listed as source ID 2246133514871244672.

The object has also been observed by the {\it Transiting Exoplanet Survey Satellite} (\tess, \citealt{tess}) in sectors 15, 16, 17, 18, 24, and 25, with 2- and 30-minute cadences. We performed a timing analysis of the data obtained in sector 15 with a 2-minute cadence. 
Figure~\ref{fig:tess2045} shows the resulting light curve and a Lomb--Scargle periodogram that is based on it. 
The periodogram reveals a $2.979 \pm 0.006$~-hour periodic signal (and its harmonics) as well as a strong and complex variability. The derived period is typical of the orbital periods of CVs (e.g., \citealt{Warner}). In the variability power spectrum, we can also distinguish a break of the continuum at $\sim 10^{-3}$~Hz, which is also typical of CVs \citep{mikej2010,2015SciA....1E0686S}.

The optical spectrum (Fig.~\ref{fig:spec_lc}) of the object reveals emission lines of the Balmer series from H$_\alpha$ to H$_\delta$ as well as emission lines of helium: HeI$\lambda 4922$, HeI$\lambda 6678$, and HeI$\lambda 7065$. The CIII + NIII$\lambda\lambda 4640,\,4650$ Bowen blend is also visible (see Fig.~\ref{Bowen:blend}), but the HeII$\lambda 4686$ and HeI$\lambda 5876$ lines are not significantly detected. Overall, these spectral features are characteristic of CVs, especially of polars \citep{Warner}. The Balmer decrement ${\rm H}_{\alpha}/{\rm H}_{\beta} = 3.5 \pm 0.9$. This corresponds to a reddening of $E(B-V) = 0.17 \pm 0.22$, which is consistent with the interstellar extinction to the source.

During the 146 minutes of AZT-33IK photometry, we obtained 540 images with an exposure of 15~s. Over this period, the reference star magnitude varied by $0.005^m$ (rms), while \stwo\ exhibited much stronger variations (see Fig.~\ref{fig:spec_lc}). The behavior of this short-term light curve is consistent with the 2.98-hour orbital modulation unveiled in the \tess\ long-term light curve. 

\subsubsection{\sthree}

This X-ray source is identified with a nonperiodic strongly variable star\footnote{https://asas-sn.osu.edu/variables/468427} from the ASAS-SN catalog of variable stars I \citep{asas18}, with $V_{mean}=16.46$ and $\Delta V=2.11$. In this catalog, the object is classified as a young stellar object with a classification probability of 0.528. In the \gaia\ EDR3 catalog, the object is listed as source ID 2225473558247025152. 

The optical spectrum of the object (Fig.~\ref{fig:spec_lc}) shows H$_{\alpha}$ and HeI$\lambda 6678$ emission lines, a weak H$_{\beta}$ emission line, and absorption lines of Na ($\lambda 5890$) and O$_2$. Only a lower limit can be placed on the Balmer decrement: ${\rm H}_{\alpha}/{\rm H}_{\beta} > 2.1$. The absence of emission lines in the short-wave part of the spectrum is likely due to the much higher extinction toward \sthree\ compared to \sone\ and \stwo. 

During the 127 minutes of AZT-33IK photometry, we obtained 240 images with an exposure of 30~s. The last 30 minutes of observations were made at dusk, at a Sun altitude of ($-12^{\circ}$, $-16^{\circ}$) degrees. As a result, the magnitudes measured after 90 minutes from the start have a lower S/N. The light curve of the object (see Fig.~\ref{fig:spec_lc}) exhibits modest variations, which are, however, much greater than the deviations in the light curve of the reference star ($0.011^m$ rms). It is impossible to draw any conclusions on an orbital or other periodicity based on this short-term light curve.

Based on the absolute magnitude $M_V \sim 4.8$, the luminosity of the object is too low for Be stars, which are also characterized by emission lines of the Balmer series. Furthermore, the variability of the object is too high for Be stars, $\Delta V=\pm 2.1$. Together with the X-ray luminosity and spectrum, this strongly indicates that \sthree\ is a CV.  

\section{Discussion}

We can use the measured X-ray fluxes, \gaia\ distances, and absolute optical magnitudes to estimate the X-ray and optical luminosities of the objects. This information and that on the equivalent width ($EW$) of the $H_{\beta}$ line in the optical spectra is given in Table~\ref{tab:prop}. For \sthree, only an upper limit on $EW$ ($H_{\beta}$) is available.

The measured X-ray luminosities are typical of CVs (e.g., \citealt{Sazonov_2006,Mukai_2017}). The fairly high $L_X/L_V$ ratios ($\sim 0.1$ -- 0.5) found for the objects indicate moderate accretion rates ($\dot{M}\sim {\rm a~few}\times 10^{15}$ -- ${\rm a~few}\times 10^{17}$~g~s$^{-1}$) onto the white dwarf \citep{patterson}\footnote{These authors used the 0.2--4~keV energy band, while we use 2--10~keV; however, the ratio of the fluxes in these bands is close to unity, $F(2-10)/F(0.2-4)\sim 1$ -- 1.5, for optically thin emission with $kT\gtrsim 5$~keV, as is the case for the objects under consideration.}. This appears to be consistent with their fairly high X-ray luminosities, taking into account the large intrinsic scatter in the $L_X/L_V$ versus $\dot{M}$ and $L_X$ versus $\dot{M}$ relations for CVs (see figs. 6 and 7 in \citealt{patterson}) and the substantial optical variability of the objects studied here. The obtained values of $L_X/L_V$ and $EW$ (${\rm H}_{\beta}$) similarly agree satisfactorily with the known correlation of these quantities for CVs \citep{patterson}.

With $L_X\sim 3\times 10^{33}$~erg~s$^{-1}$ (2--10~keV), it is highly likely that \sone\ is an intermediate polar. \stwo\ and \sthree, both having $L_X\sim 2\times 10^{32}$~erg~s$^{-1}$, are also likely to be magnetic CVs, but their luminosities can also pertain to nonmagnetic CVs (dwarf novae). The optical spectrum of \stwo\ includes a CIII + NIII$\lambda\lambda 4640,\,4650$ Bowen blend, the flux in which is comparable to the flux in H$_\beta$. This suggests that the object may be a polar \citep{Schachter_1991,Warner}, although this spectral feature can also be observed in intermediate polars (e.g., \citealt{Harlaftis_1999}). Furthermore, the bright absolute magnitude, $M_V = 3.3$, and the moderate optical variability, $\Delta V = \pm 0.3$, are not typical of dwarf novae \citep{Warner}. The same is true for \sone, for which $M_V = 2.3$ and $\Delta V=\pm 0.2$.

\begin{table*}
  \caption{Physical properties of the CVs} 
  \label{tab:prop}
  \centering
  \begin{tabular}{lcccc}
    \noalign{\doubleline}   
    Object & $L_V$, erg~s$^{-1}$ & $L_X$, erg~s$^{-1}$ & $\log(L_X/L_V)$ & $EW ({\rm H}_{\beta}$), \AA\\
    \noalign{\hrule}
    \sone & $(4.8 \pm 0.9)\times 10^{33}$ & $(2.7 \pm0.2)~\times 10^{33}$ & $-0.26 \pm 0.09$ & $8.2\pm 0.6$ \\
    \stwo & $(1.9 \pm 0.5) \times 10^{33}$ & $(2.6 \pm0.2)~\times 10^{32}$ & $-0.87 \pm 0.12$ & $1.7\pm 0.5$\\
    \sthree & $0.5^{+2.9}_{-0.4} \times 10^{33}$ & $(1.6 \pm 0.2)~\times10^{32}$ & $-0.49 \pm 0.84$ & $<3.2$\\
    \noalign{\hrule}
  \end{tabular}
  \tablefoot{
  $L_X$ is the X-ray luminosity in the 2--10 keV energy band, based on the \ero\ and \art\ data and the best-fit power-law spectral models from Table~\ref{tab:xray_table}, corrected for the Galactic absorption.
  }
\end{table*}

\section{Conclusion}

We reported the discovery of three CV candidates in the 4--12 keV \art\ source catalog obtained after the first  year of the \srg\ all-sky survey \citep{Pavlinsky_2021_cat}. These sources have also been detected by the \ero\ telescope on board the \srg. All three objects were previously known as X-ray sources from the \rosat\ all-sky survey and \xmm\ slew survey, but their nature has remained unknown. 

The X-ray spectra obtained by \ero\ and \art\ in the 0.2--20~keV energy range for \sone\ and in the 0.2--12~keV band for \stwo\ and \sthree\ are consistent with optically thin thermal emission with a temperature $kT\gtrsim 18$~keV for \sone\ and \sthree\ and $kT\gtrsim 5$~keV for \stwo. Together with the inferred high X-ray luminosities ($2\times 10^{32}$ -- $3\times 10^{33}$~erg~s$^{-1}$), this strongly suggests that all three sources are CVs. 

We have conducted optical photometry and spectroscopy of these objects using the AZT-33IK 1.6 m telescope of the Sayan Observatory. The optical properties confirm the CV nature of the objects. We conclude that \sone\ is an intermediate polar, \stwo\  likely is a polar or intermediate polar, and \sthree\ may be a magnetic or nonmagnetic CV. We also measured an orbital period of 2.98~hours for \stwo based on archival \tess\ data. 

This research is a continuation of the work of optically identifying X-ray sources \citep{Zaznobin_2021} that were detected during the \art\ all-sky survey. Three out of the planned eight \srg\ all-sky surveys have now been completed. We expect to find many new CVs during this survey and will continue our optical follow-up program. 

\begin{acknowledgements}

We appreciate helpful suggestions from the referee.
The measurements with the AZT-33IK telescope were performed within the basic financing of the FNI II.16 program and were obtained using the equipment of the Angara sharing center\footnote{http://ckp-rf.ru/ckp/3056/}. In this study, we used observational data from the \ero\ and \art\ telescopes on board \srg. The \srg\ observatory was built by Roskosmos in the interests of the Russian Academy of Sciences represented by its Space Research Institute (IKI) within the framework of the Russian Federal Space Program, with the participation of the Deutsches Zentrum f\"{u}r Luft- und Raumfahrt (DLR). The \srg/\ero\ X-ray telescope was built by a consortium of German Institutes led by MPE, and supported by DLR. The \srg\ spacecraft was designed, built, launched, and is operated by the Lavochkin Association and its subcontractors. The science data are downlinked via the Deep Space Network Antennae in Bear Lakes, Ussurijsk, and Baykonur, funded by Roskosmos. The \ero\ data used in this work were processed using the \emph{eSASS} software system developed by the German \ero\ consortium and the proprietary data reduction and analysis software developed by the Russian \ero\ Consortium. This work was supported by grant no. 21-12-00210 from the Russian Science Foundation. 

\end{acknowledgements}

\bibliographystyle{aa} 
\bibliography{references.bib}
 

\end{document}